\documentclass[sigconf]{acmart}
\usepackage{amsmath}
\usepackage{mathbbol}
\usepackage{booktabs}
\usepackage{soul}
\usepackage{url}
\usepackage{scalerel}
\usepackage{balance}
\usepackage{cleveref}

\usepackage[linesnumbered,ruled,vlined]{algorithm2e}
\usepackage{algpseudocode}

\crefformat{section}{\S#2#1#3} 
\crefformat{subsection}{\S#2#1#3}
\crefformat{subsubsection}{\S#2#1#3}

\author{Junhan Yang}
\affiliation{%
  \institution{University of Science and Technology of China}}
\email{yangjun2@mail.ustc.edu.cn}

\author{Zheng Liu}
\affiliation{%
  \institution{Microsoft Research Asia}}
\email{zhengliu@microsoft.com}

\author{Bowen Jin}
\affiliation{%
  \institution{Tsinghua University}}
\email{jbw17@mails.tsinghua.edu.cn}

\author{Jianxun Lian}
\affiliation{%
  \institution{Microsoft Research Asia}}
\email{jianxun.lian@microsoft.com}

\author{Defu Lian}
\affiliation{%
  \institution{University of Science and Technology of China}}
\email{liandefu@ustc.edu.cn}

\author{Akshay Soni}
\affiliation{%
  \institution{Microsoft}}
\email{akshay.soni@microsoft.com}

\author{Eun Yong Kang}
\affiliation{%
  \institution{Microsoft}}
\email{eun.kang@microsoft.com}

\author{Yajun Wang}
\affiliation{%
  \institution{Microsoft}}
\email{	yajunw@microsoft.com}
\author{Guangzhong Sun}
\affiliation{%
  \institution{University of Science and Technology of China}}
\email{gzsun@ustc.edu.cn}
\author{Xing Xie}
\affiliation{%
  \institution{Microsoft Research Asia}}
\email{xingx@microsoft.com}

\begin{document}

\title{Hybrid Encoder: Towards Efficient and Precise Native Ads Recommendation via Hybrid Transformer Encoding Networks}

\begin{abstract}
Transformer encoding networks have been proved to be a powerful tool of understanding natural languages. They are playing a critical role in native ads service, which facilitates the recommendation of appropriate ads based on user's web browsing history. For the sake of efficient recommendation, conventional methods would generate user and advertisement embeddings independently with a siamese transformer encoder, such that approximate nearest neighbour search (ANN) can be leveraged. Given that the underlying semantic about user and ad can be complicated, such independently generated embeddings are prone to information loss, which leads to inferior recommendation quality. Although another encoding strategy, the cross encoder, can be much more accurate, it will lead to huge running cost and become infeasible for realtime services, like native ads recommendation.

In this work, we propose hybrid encoder, which makes efficient and precise native ads recommendation through two consecutive steps: retrieval and ranking. In the retrieval step, user and ad are encoded with a siamese component, which enables relevant candidates to be retrieved via ANN search. In the ranking step, it further represents each ad with disentangled embeddings and each user with ad-related embeddings, which contributes to the fine-grained selection of high-quality ads from the candidate set. Both steps are light-weighted, thanks to the pre-computed and cached intermedia results. To optimize the hybrid encoder's performance in this two-stage workflow, a progressive training pipeline is developed, which builds up the model's capability in the retrieval and ranking task step-by-step. The hybrid encoder's effectiveness is experimentally verified: with very little additional cost, it outperforms the siamese encoder significantly and achieves comparable recommendation quality as the cross encoder. 
\end{abstract}


\maketitle

\section{Introduction}
\begin{figure}[t]
\centering
\includegraphics[width=0.92\linewidth]{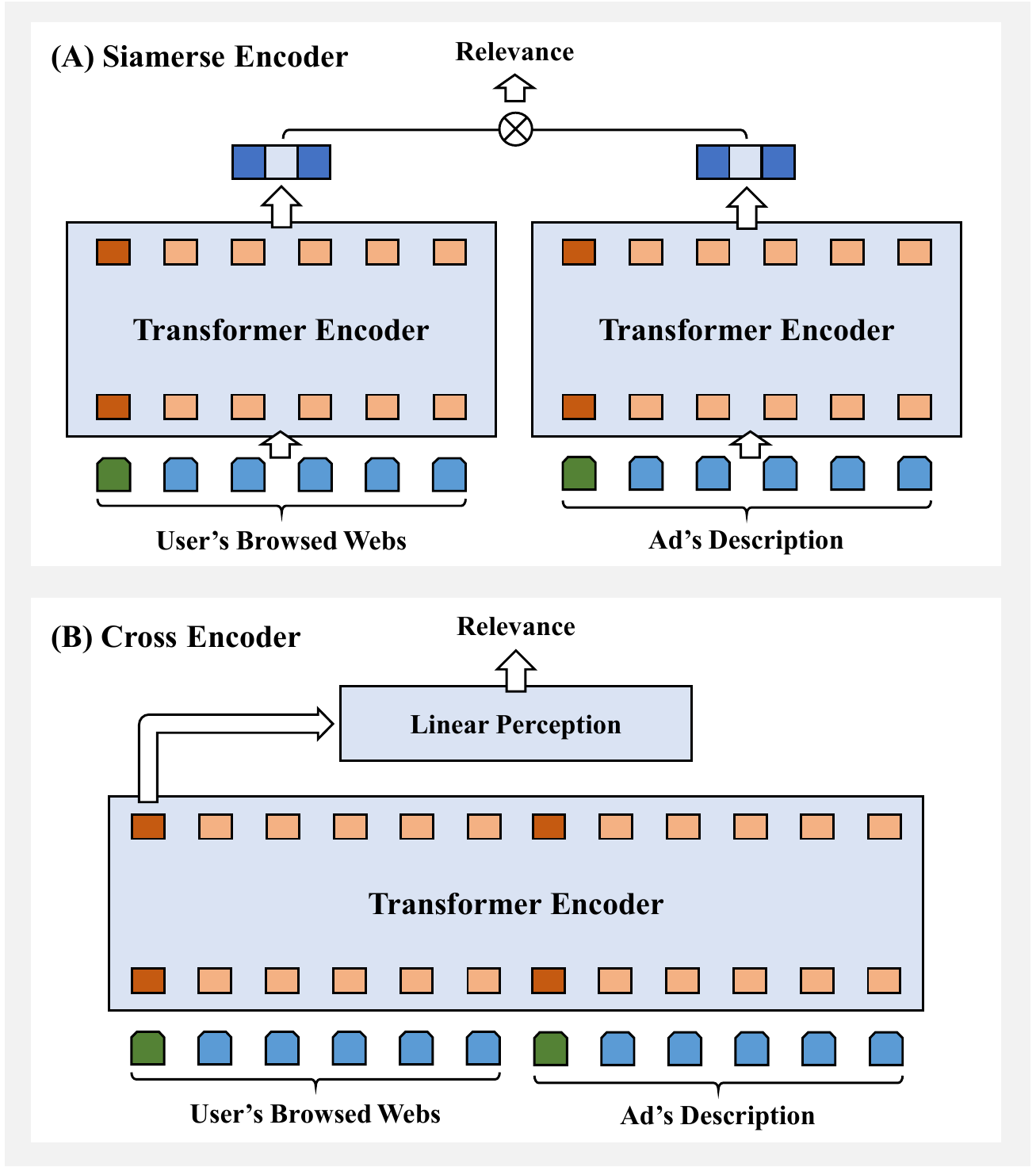}
\caption{Illustration of conventional transformer encoders. (A) siamese encoder: user's browsed webs and ad's description are encoded independently, whose relevance is calculated as the embedding similarity. (B) cross encoder: user's browsed webs and ad's description are concatenated into one token list and jointly encoded by a unified transformer; the relevance is computed based on final hidden-states, usually the one corresponding to [CLS] token. }
\label{fig:baseline}
\end{figure}

Native ads are important ways of online advertising, where advertisements are
imperceptibly injected into the webpages being browsed by the users. In contrast to the conventional search ads where queries are explicitly specified by user, native ads are recommended implicitly based on user's web-browsing history. Apparently, high-quality recommendation calls for deep understanding of user behavior and ad content. In recent years, transformer-based models turn out to be popular solutions due to their superior capability on natural language processing, e.g., Bert \cite{devlin2018bert} and RoBerta \cite{liu2019roberta}. The existing methods would represent a pair of input sequences independently with a siamese encoder \cite{reimers2019sentence,lu2020twinbert} (as the two-tower architecture in Figure \ref{fig:baseline}.A), and measure their relevance based on the embeddings' similarity, such as cosine and inner-product. With these approaches, relevant ads can be efficiently acquired via ANN search, like PQ \cite{jegou2010product, ge2013optimized} and HNWS \cite{malkov2018efficient}. However, one obvious defect is that there is no interaction between the user and ad encoding process. Given that the underlying semantic about user and ad can be complicated, such independently generated embeddings is prone to information loss, thus cannot reflect the user-ad relationship precisely. In contrast, another form of transformer encoding networks, the cross-encoder (as the one-tower structure in Figure \ref{fig:baseline}.B), can be much more accurate \cite{luan2020sparse,lee2019latent}, where the input sequences may fully attend to each other during the encoding process. Unfortunately, the cross-encoder is highly time-consuming and hardly feasible 
for realtime services like native ads recommendation, as every pair of user and ad needs to be encoded and compared from scratch. The above challenges bring us to a challenging problem: whether some sort of encoding architecture can be both efficient and precise. Motivated by this challenge, we propose hybrid encoder, which addresses the above problem with the following featurized designs.

$\bullet$ \textbf{Two-stage workflow}. Instead of making recommendation in one shot, hybrid encoder selects appropriate ads through two consecutive steps: retrieval and ranking, with two sets of user and ad representations. First of all, it represents user and ad independently with a siamese encoder, such that relevant candidates can be retrieved via ANN search. Secondly, it generates disentangled ad embeddings and ad-related user embeddings, based on which high-quality ads can be identified from the candidate set. Both steps are highly efficient, which makes the time cost affordable in practice.

$\bullet$ \textbf{Fast ranking}. Unlike the conventional time-consuming cross-encoder, hybrid-encoder establishes the user-ad interaction by intensively re-using the cached hidden-states and pre-computation result. On the one hand, it caches the transformer's last-layer hidden-states for the user, which will comprehensively preserve the information about user's web-browsing behaviors. On the other hand, it pre-computes the disentangled embeddings for each ad. While an ad is chosen as a user's candidate, the disentangled ad embeddings will attend to the user's cached hidden-states, which gives rise to the ``ad-related'' user embeddings. With such embeddings, necessary information for analyzing user-ad relationship can be effectively extracted, which in return contributes to the selection of best ads from the candidate set. Apparently, most of the involved operations are light-weighted, e.g., inner-product and summation, whose time cost is almost neglectable compared with the feed-forward inference of transformer. Therefore, the running time of the ranking step is approximately up-bounded by one pass of transformer encoding.

$\bullet$ \textbf{Progressive Training Pipeline}. Although both the retrieval and ranking step aim to find out appropriate advertisements w.r.t. the presented user, they are actually deployed for distinct purposes, thus need to be learned in different ways. In this work, a progressive learning strategy is adopted to optimize the overall performance of the two-stage workflow. For one thing, the retrieval step is to select coarse-grained candidates from the whole advertisement corpus; therefore, it is trained with \textit{global contrast learning}, which discriminates the ground-truth from the negative ones sampled from the global scope. For another thing, the ranking step is to select the best advertisements from the candidate set generated by a certain retrieval policy. Therefore, the model is further trained with the \textit{local contrast learning}, where a set of candidate advertisements are sampled based on the retrieval policy learned in the first stage, and the model is required to identify the ground-truth given such ``hard negative samples''.

Comprehensive empirical investigation are performed on Microsoft Audience Ads\footnote{https://about.ads.microsoft.com/en-us/solutions/microsoft-audience-network/microsoft-audience-ads} platform, where the hybrid encoder's effectiveness is verified from the following aspects. \textbf{I.} It achieves comparable recommendation quality as the accurate but expensive cross-encoder, given that the running time is sufficiently large. \textbf{II.} It significantly outperforms the conventional siamese encoder when the time cost is limited within a tolerable range as required by the production. Additional studies are also made with public-available dataset, which further confirms the above findings. 

To sum up, the following contributions are made in this work.
\\
$\bullet$ We propose hybrid encoder, which recommends native ads with a two-stage workflow: the retrieval step collects coarse-grained candidates from the whole corpus; the ranking step selects the best ads from the given candidates. Both steps are highly efficient, which collaboratively produce precise recommendation for the user. 
\\
$\bullet$ The proposed model is trained progressively through global contrast learning and local contrast learning, which optimizes the overall performance of the two-stage recommendation workflow. 
\\
$\bullet$ The effectiveness of hybrid encoder is verified with large-scale data from native ads production. Compared with conventional transformer encoders, hybrid encoder is competitive in terms of both recommendation precision and working efficiency.


\section{Related Work}
In this section, related works are discussed from the following perspectives: 1) native ads recommendation, and 2) transformer encoding networks.

\subsection{Native Ads Recommendation}
Native ads is a relatively new form of online advertising, where advertisements are recommended to user based on their web-browsing behaviors. A typical recommendation workflow is composed of the following operations. Particularly, each advertisement is characterized with a compact description, e.g., new apple ipad mini. When an advertising request is presented, it turns user's browsed web-pages into ``implicit queries'' and finds out the most relevant ads-descriptions. Finally, the selected ads will be displayed imperceptibly on the web-pages being browsed by the user. In the early days, the recommendation process would intensively make use of the retrieval algorithms built on explicit features: each ads-description is further split into a set of keywords, based on which an inverted index can be built to organize the whole advertisements. To make the index more accurate and compact, conventional IR techniques like TF-IDF and BM25 are adopted so as to identify the most important keywords for index \cite{robertson2009probabilistic, manning2008introduction}. Given an implicit query from user, the system will extract its keywords in the first place, and then join with the inverted index so as to collect the advertisements within the matched entries. 

Apparently, the old-fashioned approaches can be severely limited by the ratio of keyword-overlap. To overcome such a defect, more and more effort has been made on latent retrieval \cite{fan2019mobius,huang2020embedding}, where the semantic relationship between user and ad can be measured more precisely with compact embeddings. Early representative works along this direction includes DSSM \& CDSSM \cite{huang2013learning,shen2014latent}, where multi-layer-perception and convolutional neural networks are adopted; in recent years, much more advanced approaches are continuously developed, especially based on transformer encoders and pretrained language models, such as \cite{reimers2019sentence,lu2020twinbert,karpukhin2020dense,guu2020realm,lee2019latent}.
 
\subsection{Transformer Encoding Networks}
Transformer is a powerful network structure for text encoding \cite{vaswani2017attention}, as long range interactions can be established for the input sequence on top of multi-head self-attention. In recent years, multi-layer transformer encoders have become the foundation of almost every large-scale pretrained language model, like BERT \cite{devlin2018bert} and RoBerta \cite{liu2019roberta}, with which the underlying semantic can be captured in-depth for the presented text. Given its superior capability in various NLP applications, transformer is also used as the backbone network structure for text retrieval \cite{reimers2019sentence,lu2020twinbert,karpukhin2020dense}. Typically, a two-tower transformer encoding network (also known as \textit{siamese encoder} or \textit{bi-encoder}) are deployed, which generates the latent representations for the source and target text segments; after that, the source and target relevance can be measured with the embedding similarity, e.g., cosine or inner-product. The siamese encoder naturally fits the need of large-scale retrieval/recommendation, as relevant targets can be efficiently acquired for the given source via those well-known ANN paradigms, e.g., PQ \cite{jegou2010product,ge2013optimized} and HNSW \cite{malkov2018efficient}. However, one obvious limitation about the siamese encoder is that the source and target are encoded independently; in other words, there is no interaction between the source and target's encoding process. Such a problem may lead to the ignoring of important information for the generated embeddings, which harms the model's accuracy. According to \cite{luan2020sparse}, this problem will be more severe as the input sequence becomes longer.

An intuitive way of addressing the above problem is to make use of the one-tower transformer encoder \cite{luan2020sparse}, usually referred as \textit{cross-encoder}, where the inputs are combined into one sequence and jointly processed by a unified encoding networks. However, this architecture needs to encode every pair of inputs from scratch for comparison, whose running time is prohibitive for most of the realtime services, like native ads recommendation (where hundreds or up-to-thousand comparisons need to be made within millisecond-scale). Recently, it becomes more and more emphasized on how to acquire accurate matching result in an temporally-efficient manner. One promising option is to establish the interaction between inputs on top of the transformer's hidden-states \cite{humeau2019poly,luan2020sparse}. For example, in \cite{humeau2019poly}, the hidden-states of the source sequence are mapped into a total of $K$ embeddings called context vectors; the context vectors will be further aggregated w.r.t. the target embedding via attentive pooling; finally, the aggregation result will be compared with the target embedding so as to reflect the source-target relevance. Our work also takes advantage of the hidden-states. Unlike the conventional methods, the interaction between inputs is established based on raw hidden-states of the source and disentangled embeddings of the target, which turns out to be both efficient and more effective.

\section{Methodology}

\subsection{Preliminaries}\label{sec:met-pre}
Transformer encoding networks have been proved to be a powerful tool for natural language understanding, which significantly benefit downstream tasks, like recommendation and information retrieval. Such networks consist of multiple layers of transformer blocks \cite{vaswani2017attention}, which are built upon multi-head self-attention (MHSA). For each particular layer, the input embeddings (used as $Q$, $K$ and $V$) are interacted with each other via the following operations:
\begin{equation}
\begin{gathered}
\mathrm{MHSA}(Q, K, V) = \mathrm{Concat}(head_1,...,head_h)W^O, \\
head_i = \mathrm{Attention}(QW_i^Q, KW_i^K, VW_i^V).
\end{gathered}
\end{equation}
where $W^O$, $W_i^Q$, $W_i^K$, $W_i^V$ are the projection matrices to be learned. In transformers, the scaled dot-product attention is adopted, which computes the weighted-sum of $V$ with the following equation:
\begin{equation}
    \mathrm{Attention}(Q, K, V) = \mathrm{softmax}(\frac{QK^T}{\sqrt{d}})V,
\end{equation}
where $d$ indicates the value of hidden vector's dimension. Apparently, MHSA has the advantage of introducing long-range interaction for the input sequence, which brings about stronger context-awareness for the generated representations. 

While applied for IR/recommendation-related tasks, transformer encoding networks usually adopt the following two forms: the siamese encoder and the cross encoder. In the following part, we will demonstrate how the above forms of encoders are applied for the native ads recommendation scenario.

\subsubsection{siamese Encoder} The siamese encoder leverages a pair of identical sub-networks to process the user history and ads description, as shown in Figure \ref{fig:baseline} (A). Each sub-network will aggregate the last layer's hidden states for the representation of its input sequence; as a result, two embeddings $E_{u}$ and $E_{a}$ will be generated, which are of the same hidden dimension. The user-ad relevance is measured based on the similarity between the embeddings:
\begin{equation}
    Rel(u, a) = \langle E_{u}, E_{a} \rangle,
\end{equation}
where $\langle \cdot \rangle$ stands for the similarity measurement, e.g., inner-product. The good thing about siamese encoder is that the user history and ad description can be encoded once and used for relevance computation everywhere. More importantly, the generated embeddings can be indexed for ANN-search (e.g., MIPS \cite{jegou2010product}), which naturally fits the need of large-scale retrieval/recommendation. However, the siamese encoder is prone to information loss, as user and ad are independently encoded. Therefore, it is usually limited by its inferior accuracy. 

\begin{figure*}[t]
\centering
\includegraphics[width=0.85\textwidth]{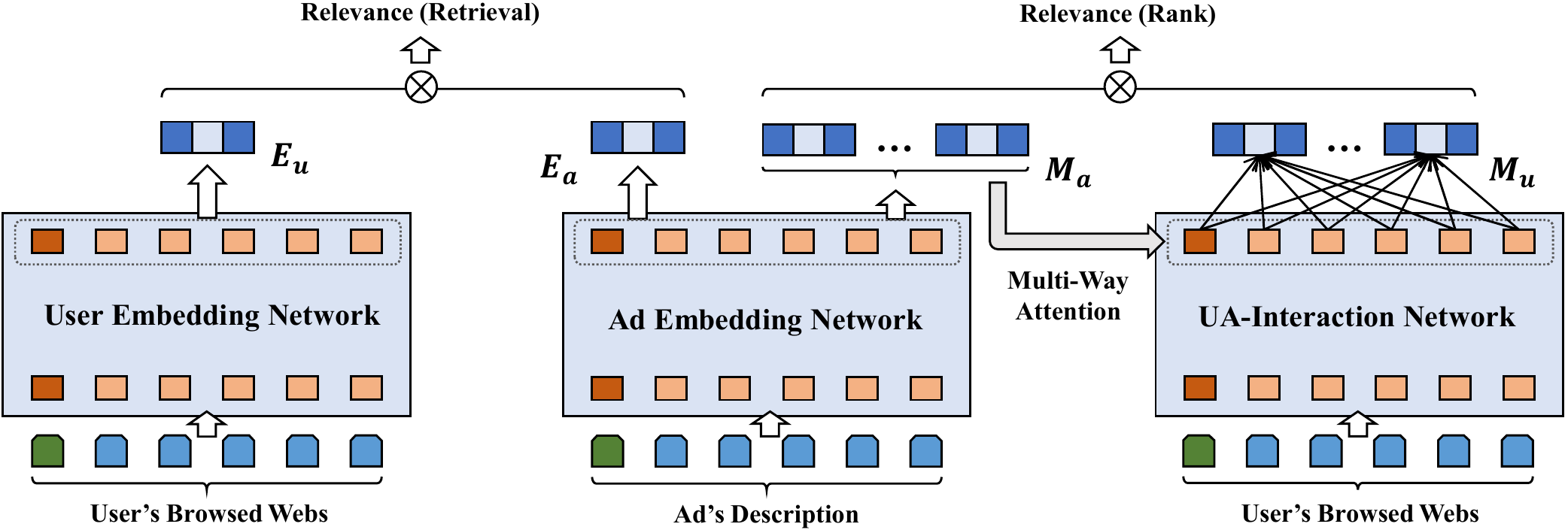}
\caption{Infrastructure of Hybrid Encoder. In the middle and left side, the user embedding $E_u$ and ad embedding $E_a$ are generated independently; both embeddings are used for the ANN search in the retrieval step. On the middle and right side, the disentangled ad embeddings $M_a$ are offline generated, which are used to attend user's cached hidden-states for the ad-related user embeddings; both embeddings are flattened into 1-D vecs and compared to measure the fine-grained user-ad relationship.}
\label{fig:hybrid}
\end{figure*}

\subsubsection{Cross Encoder} The cross encoder makes use of a unified transformer to analyze the relevance between the user and ad. In particular, the input sequences are concatenated into the following joint string marked with [CLS] and [SEP]:
\begin{equation}
    [CLS]t_0^{user},...,t_M^{user}[SEP]t_0^{ad},...,t_N^{ad},
\end{equation}
where $t_{\cdot}^{user}$ and $t_{\cdot}^{ad}$ indicate tokens from user's browsed web-pages and ad description, respectively. The joint string will be encoded by the transformer. And based on the last layer's hidden-state corresponding to [CLS], denoted as $H_{[CLS]}$, the web-ad relevance is calculated as:
\begin{equation}\label{eq:cross}
    Rel(user, ads) = \sigma(H_{[CLS]} W_A),
\end{equation}
where $W_A$ is the $d\times1$ linear projection vector for the output logit. The cross encoder is of much higher model capacity compared with the siamese encoder, as user and ad may fully get attended by each other within the encoding process. Therefore, cross encoder can be much more accurate, as important information is preserved more effectively. However, the cross encoder is highly limited in working efficiency, as every pair of user and ad need to be encoded from scratch so as to find out the most relevant result. 

\subsection{Overview of Hybrid Encoder}

\subsubsection{Problem Formulation}
We consider the scenario where natives ads recommended are based on user's web-browsing behaviors. In this place, the following definition is presented, which clarifies the objective of native ads recommendation (NAR). 
\begin{definition}
(NAR) Given user's web-browsing history, NAR looks for the advertisement $a$, which gives rise to the maximum probability of being clicked.
\end{definition}
It is apparent that the NAR problem can be quantitatively expressed by the following optimization problem: 
\begin{equation}
   \max \mathbb{E}_{(u,a) \sim P(u,a)} \log \pi(a|u),
\end{equation}
where $P(u,a)$ stands for the joint probability of user-$u$-click-ad-$a$, and $\pi(a|u)$ refers to the advertisement's probability of being recommended by NAR conditioned on the user's web-browsing history. By optimizing the above objective, the recommendation algorithm is expected to produce the maximum amount of ad-clicks, which contributes to the revenue growth of the whole market. 

\subsubsection{Two-Stage Processing}\label{sec:met-ov-two} To facilitate efficient and precise native ads recommendation, a two-stage workflow is employed, which contains the consecutive steps of retrieval and ranking. The retrieval step is to select relevant candidate ads w.r.t. the presented user. Given that the size of the whole advertisement set can be huge (usually as large as a few millions), the retrieval steps must be highly efficient. Thus, approximate nearest neighbour search turns out to be a natural choice. To make use of it, user history and ad description are independently encoded as $E_{u}$ and $E_{a}$, and the user-ad relevance is measured with the embedding similarity. As a result, we'll efficiently get the candidate ads for the retrieval step via ANN search:
\begin{equation}
    \Omega_{retr} = \mathrm{ANN}(E_u, \{E_a|\Omega_A\})
\end{equation}
where $\{E_a|\Omega_A\}$ denotes the embeddings of whole ads. Following the retrieval step, the ranking step is carried out so as to select the best set of ads out of the coarse-grained candidates. For the sake of better recommendation quality, a ranking function $f(\cdot)$ will be utilized, which is able to establish the interaction between user and each of the candidate ads. Finally, the fine-grained set of ads for recommendation can be determined based on the ranking scores:
\begin{equation}
    \Omega_{rank} = \mathrm{Top}\text{-}K \{ f(u,a) | \Omega_{retr} \}
\end{equation}

\subsubsection{Online Cost Analysis}\label{sec:met-ov-time} The running time of the above two-stage processing is constituted of three parts: the user and ad embedding generation, the ANN search, and the ranking of candidates. However, because the in-market ads are temporally invariant and given beforehand, the ad embeddings can all be generated offline; besides, compared with the model's inference cost, the ANN search is fast and its time complexity is constant (or sub-linear). Therefore, the online cost will be determined mainly by two factors: the \textit{generation of user embedding} and \textit{the ranking of candidates}. The generation of user embedding takes one-pass of transformer encoding. And with the conventional cross-encoder, it takes  $|\Omega_{retr}|$ rounds of transformer encoding to rank the candidates. Finally, there will be $|\Omega_{retr}|+1$ rounds of transformer encoding in total, which is still temporally infeasible for native ads production. Apparently, the ranking step turns out to be the efficiency bottleneck of the two-stage workflow. In the following discussion, a significantly accelerated ranking operation will be introduced, where the time cost is reduced to $\epsilon$ transformer encoding ($\epsilon\leq1$). As a result, it will merely take $1+\epsilon$ transformer encoding, which makes hybrid encoder competitive in working efficiency.

\subsection{Model Architecture}
The architecture of hybrid encoder is shown in Figure \ref{fig:hybrid}, where multiple transformers are jointly deployed for the hybrid encoding process. On the left and middle side of the model, the deployed transformers take the form of a siamese encoder, where user and ad embeddings, $E_u$ and $E_a$, are generated independently. These embeddings are used for the ANN search of the retrieval step. On the middle and right side, the disentangled ad embeddings $M_a$ are generated, which may capture ad semantic more comprehensively; meanwhile, the disentangled ad embeddings will attend to the cached hidden-states of the ua-interaction network, which gives rise to a total of $|M_a|$ ``ad-related'' user embeddings $M_u$. Finally, the disentangled ad embeddings and ad-related user embeddings are flatten into 1-D vectors and compared, which helps to make fine-grained analysis of user-ad relationship for the ranking step.

\subsubsection{User and Ad Embedding Generation}\label{sec:met-mod-sia}
The user's browsed web-pages and ads' descriptions are independently feed into the user embedding network and the ad embedding network. Unlike the default practice where the [CLS] token's hidden-state in the last-layer is taken as the representation result, we use the mean-pooling of all tokens' hidden-states in the last-layer as the user and ad embeddings, which is suggested to be more effective \cite{reimers2019sentence,lu2020twinbert}:
\begin{equation}
\begin{gathered}
    E_u = \text{mean}(H^u), ~where~ H^u = \text{U-Net}(\text{user's browsed webs}); \\
    E_a = \text{mean}(H^a), ~where~ H^a = \text{A-Net}(\text{ad description}).    
\end{gathered}
\end{equation}
As the user and ad embeddings need to support the ANN search operation in the retrieval step, $E_u$ and $E_a$ must be within the same latent space, whose similarity is directly comparable. As a result, the user-ad relevance for the retrieval step is computed as: 
\begin{equation}\label{eq:siamese}
    Rel_{retr}(u,a) = \langle E_u, E_a \rangle.
\end{equation}
In our implementation, we use inner-product as the similarity measure and adopt HNSW for the search of approximate neighbours.

\subsubsection{Fast Ranking with UA Interaction}
The interaction between user and ad is established via ua-interaction network, as shown on the right side of Figure \ref{fig:hybrid}. In this stage, the following three operations are carried out step-by-step. First of all, it generates $K$ disentangled ad embeddings $M_a$ with multi-head attention \cite{lin2017structured}: 
\begin{equation}
\begin{gathered}
    M_a : \{ M_a^i = \sum\nolimits_{H^a} \alpha^i_j H^a_j | i=1,...,K \}, \\
    ~ where ~ \boldsymbol{\alpha}^i = \text{softmax}(H^a, \boldsymbol{\theta}^i).
\end{gathered}
\end{equation}
In the above equation, $\{\boldsymbol{\theta}^i\}_1^K$ stands for the attention heads, which aggregate the hidden-states of the advertisement from different views for more comprehensive representation. Secondly, each of the disentangled ad embeddings is used to attend user's last-layer hidden-states $H^{ua}$ from the ua-interaction network. As a result,the following ad-related user embeddings $M_u$ will be generated:
\begin{equation}
\begin{gathered}
    M_u: \{ M_u^i = \sum\nolimits_{H^{ua}} \alpha^i_j H^{ua}_j | i=1,...,K \}, \\
    ~ where ~ \boldsymbol{\alpha}^i = \text{softmax}(H^{ua}, M_a^i).
\end{gathered}
\end{equation}
Finally, the disentangled ad embeddings $M_a$ and the ad-related user embeddings $M_u$ are flattened and concatenated into 1-D vectors, whose similarity is used as the measurement of user-ad relevance for the ranking step:
\begin{equation}\label{eq:hybrid}
\begin{gathered}
    Rel_{rank}(u,a) = \langle \bar{M}_u, \bar{M}_a \rangle, \\
    where ~ \bar{M}_u = \text{cat}([M_u^1,...,M_u^K]), ~ \bar{M}_a = \text{cat}([M_a^1,...,M_a^K]).
\end{gathered}
\end{equation}
$\bullet$ \textbf{Remarks}. Compared with the expensive cross-encoding process, there are two notable characters about the proposed ua-interaction: 1) the interaction is uni-directional: with only user's hidden-states attended by the representation of ad. And 2) the interaction is introduced not from the very bottom, but at the last-layer of the transformer. Despite of its simplicity, the proposed method is desirable from the following perspectives. 

First of all, it leverages the pre-computation result (i.e., the disentangled ad embeddings) and the cached hidden-states (from ua-interaction network) to establish interaction between user and ad. Therefore, it merely incurs one pass of feed-forward inference of the ua-interaction network. Besides, it's empirically found that a relatively small-scale network (compared with the user embedding network) may already be sufficient for making high-quality interaction. Therefore, it only brings about $\epsilon$ ($\epsilon\leq1$) of the transformer encoding cost as the generation of user embedding, which makes the ranking step highly competitive in time efficiency. 

Secondly, the proposed ua-interaction is more effective in preserving the underlying information of user and ad. On the one hand, the ad descriptions are usually simple and compact; even without being interacted by the user, the disentangled embeddings may still comprehensively represent the advertisement. On the other hand, despite that user history can be complicated, the necessary information for measuring user-ad relationship will be fully extracted when attended by the disentangled ad embeddings. As a result, the relevance between user and ad will be measured more precisely based on such informative embeddings, which significantly benefits the identification of the best candidates.

\begin{figure}[t]
\centering
\includegraphics[width=1.0\linewidth]{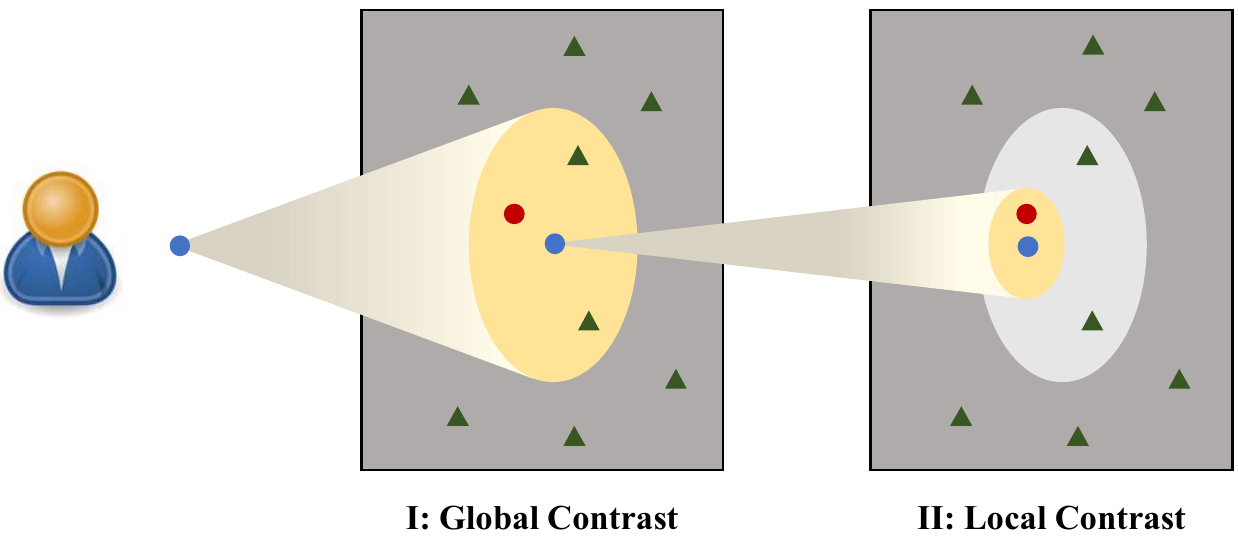}
\caption{Illustration of the progressive training pipeline, where the blue and red dots indicate the user embedding and the ground-truth ad embedding, while the green triangles stands for the irrelevant ad embeddings. In the 1st stage, the global contrast learning discriminates the ground-truth from the global negatives, which leads to a relatively coarse-grained scope of candidates (the yellow region); in the 2nd stage, the local contrast learning identifies the ground-truth within the candidate scope, which pushes the user embedding even closer to the ground-truth ad embedding.}
\label{fig:train}
\end{figure}

\subsection{Progressive Training Pipeline}\label{sec:met-pro}
Hybrid encoder is trained progressively with two consecutive steps (as Figure \ref{fig:train}). Firstly, the \textit{global contrast learning} is carried out for the user and ad embedding networks; in this step, the model learns to discriminate the ground-truth ad from the global negatives. As a result, it makes the ground-truth to be confined within the neighbourhood of user embedding, which can be retrieved via ANN-search. Secondly, the \textit{local contrast learning} is performed for the ua-interaction network, where the model continues to learn the selection of ground-truth from the presented candidates. With this operation, the user and ad representations (more specifically, the flattened ad-related user embeddings and disentangled ad embeddings) will be further pushed together, which improves the recommendation accuracy of the ranking step.

\subsubsection{Global Contrast Learning} The global contrast learning is carried out in the form of binary classification. Firstly, a positive user-ad case $(u,a)$ is sampled from the production log; then, a total of $n$ negatives ads $\{a'_i\}_n$ are sampled from the global ad corpus $A$. As a result, the following training objective is formulated:
\begin{equation}
\begin{gathered}
\mathrm{E}_{(u,a)\sim P(u,a)} \Big( \log\big(\sigma(u,a)\big) - \sum\nolimits_{\{a'_i\}_n \sim A} \log\big(\sigma(u,a'_i)\big) \Big), \\
~ where ~ \sigma(u,a) = \text{exp}\big(Rel_{retr}(u,a)\big). 
\end{gathered}
\end{equation}
With the optimization of the above objective, the user and ground-truth ad embeddings will come close towards each other; meanwhile, the global negatives' embeddings will be pushed away from the user embedding. Therefore, it enables the retrieval step to find out relevant candidate ads via ANN search.

\subsubsection{Local Contrast Learning}\label{sec:met:lcl} The local contrast learning is performed w.r.t. the well trained retrieval strategy. For each positive $(u,a)$ sample, the negative cases are sampled within the scope of candidates, i.e., the neighbour of user embedding $E_u$: $\text{N}(E_u)$, based on which the training objective is formulated as follows:
\begin{equation}
\begin{gathered}
\mathrm{E}_{(u,a)\sim P(u,a)} \Big( \log\big(\sigma(u,a)\big) - \sum\nolimits_{\{a'_i\}_n \sim \text{N}(E_u)} \log\big(\sigma(u,a'_i)\big) \Big), \\
~ where ~ \sigma(u,a) = \text{exp}\big(Rel_{rank}(u,a)\big). 
\end{gathered}
\end{equation}
Notice that $\text{N}(E_u)$ does not necessarily mean the nearest neighbours of $(E_u)$. In fact, it is empirically found that the using the nearest neighbours will lead to inferior training outcome, where the underlying reasons are two-fold: firstly the nearest neighbours may give rise to a high risk of false-negatives, i.e., the sampled ads are non-clicked but relevant to user; secondly, the nearest neighbours can be monotonous, i.e., the sampled negatives are probably associated with similar semantics, thus reducing the information which can be learned form the negative samples. In our implementation, we will firstly collect a large number of neighbours of $E_u$, and then randomly sample a handful of cases as the training negatives.

\section{Experimental Studies}

\begin{table}[t]
\centering
\includegraphics[width=1.05\linewidth]{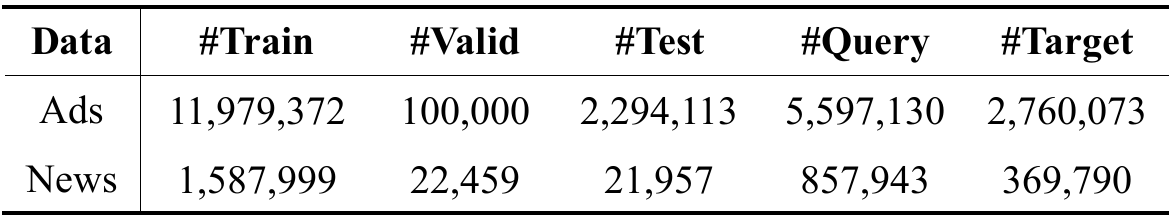}
\caption{Specs about the datasets. \#Train/Valid/Test: number of samples in train/valid/test set; \#Query: number of input queries (ads: \#user, news: \#headline); \#Target: number of recommendation targets (ads: \#ad, news: \#annotation).}
\label{tab:data}
\end{table}

\begin{table*}[t]
\centering
\includegraphics[width=0.95\textwidth]{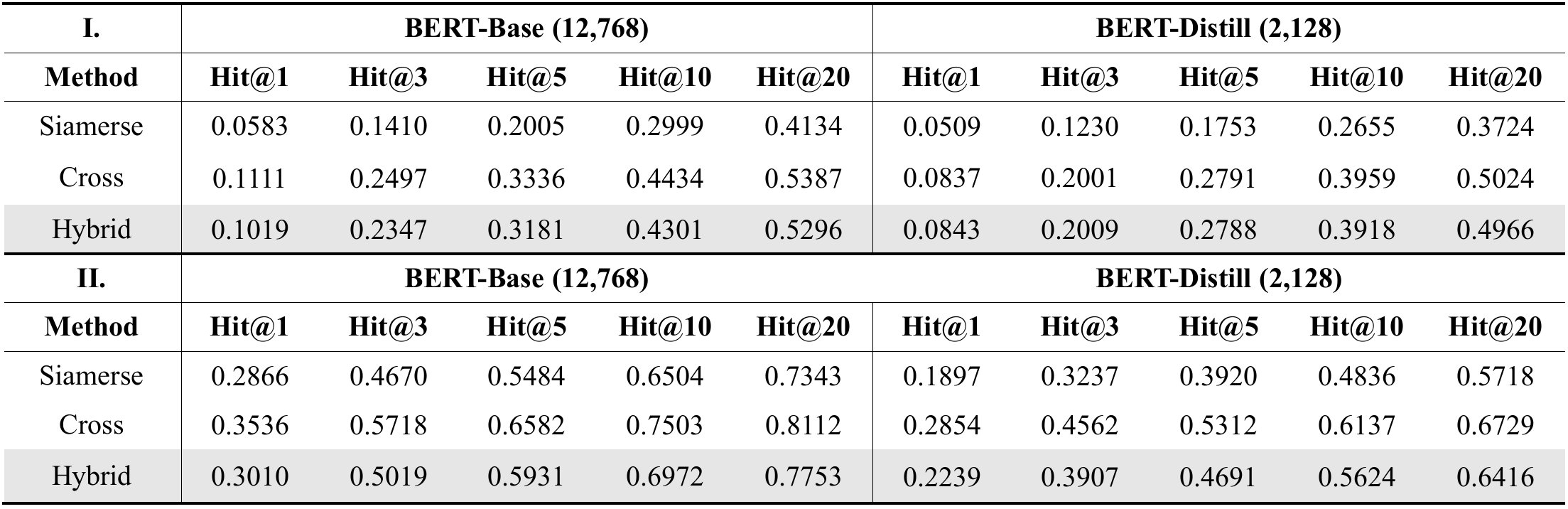}
\caption{Basic Comparison \cref{sec:exp-ana-bas}, I. \& II.: results on Native Ads and Google News, respectively. The hybrid encoder outperforms the siamese encoder on both datasets, and it generates comparable performance as the cross encoder on Native Ads dataset.}
\label{tab:basic}
\end{table*}

\begin{table*}[t]
\centering
\includegraphics[width=0.95\textwidth]{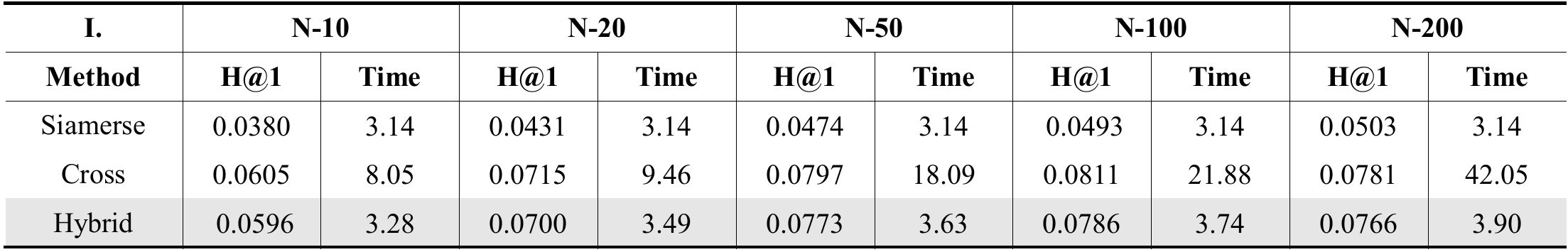}
\caption{Hit-Time Ratio Analysis \cref{sec:exp-ana-tim} (time measured in ms). The hybrid encoder achieves comparable performance as the cross encoder, but it merely incurs small additional cost compared with the siamese encoder.}
\label{tab:time}
\end{table*}

\subsection{Experiment Settings}
The settings about the experimental studies are specified as follows.

$\bullet$ \textbf{Datasets.} The experimental studies are carried out based on production data collected from Microsoft Audience Ads platform\footnote{https://about.ads.microsoft.com/en-us/solutions/microsoft-audience-network/microsoft-audience-ads}. The dataset is composed of large-scale user-ad pairs, in which users are characterized by their web-browsing behaviors, and web-pages/advertisements are associated with their textual descriptions. As discussed in the methodology, the recommendation algorithms are required to predict user's next-clicked ad based on her web-browsing behavior. The training, validation and testing sets are strictly partitioned in time 
in order to avoid data leakage: the training data is from 2020-05-01 to 2020-05-31, while the validation/testing data is from 2020-07-14 to 2020-08-11. The detailed specifications about the dataset are illustrated with Table \ref{tab:data}.

Although the proposed method is designed for native ads recommendation, it is also necessary to explore its performance on more generalized scenarios, which select semantic-close targets for the input query from a large-scale corpus. In this place, the Google News dataset\footnote{https://github.com/google-research-datasets/NewSHead} is adopted for further evaluation. This dataset contains the headlines of news articles (parsed from the provided urls) and their human annotated descriptions, e.g., news headline ``\textit{nba playoffs 2019 milwaukee bucks vs toronto raptors conference finals schedule}'' and human annotation ``\textit{NBA playoff Finals}''. In our experiment, the news headline is used as the input query, to which the related human annotation needs to be selected from the whole annotation corpus. The training and validation/testing set partition is consistent with \cite{headline2020} where the dataset was originally created.

$\bullet$ \textbf{Baselines}. As discussed, the hybrid encoder is compared with the two representative forms of transformer encoding networks \cite{luan2020sparse}. One is the \underline{siamese-encoder}, which is the conventional way of making use of transformers for native ads recommendation \cite{lu2020twinbert}; as hybrid encoder already includes a siamese component, we'll directly take that part out of the well-trained hybrid encoder for testing such that impact from inconsistent settings can be excluded. The other one is the \underline{cross-encoder}, which is the most accurate but temporally infeasible; the cross encoder is also learned through the same training workflow as the hybrid encoder's ranking component \cref{sec:met:lcl}, thus we can fairly compare the effect resulted from model architecture. The computation of user-ad relevance with both encoders is consistent with the previous introduction made in \cref{sec:met-mod-sia} and \cref{sec:met-pre}. In our experiments, we adopt \underline{BERT-base} \cite{devlin2018bert}, the 12-layer-768-dim bi-directional pretrained transformer, as our backbone structure. Besides, as BERT-base is still a too large-scale model for realtime service, another model based on a light-weighted 2-layer-128-dim \underline{distilled BERT} is trained for production usage.

$\bullet$ \textbf{Performance Evaluation}. The quality of natives ads recommendation is evaluated in terms of \underline{Hit@N}, which reflects how likely the ground-truth ad can be included within the top-N recommendation. To focus on the effect resulted from each model itself, the top-N recommendation is made based on the same two-stage workflow as discussed in \cref{sec:met-ov-two}, where a shared set of candidate ads (a total of $k$N, which equals to 100 by default) are provided from approximate nearest neighbour search. As introduced in \cref{sec:met-mod-sia}, the ANN search is performed with HNSW\footnote{https://github.com/nmslib/hnswlib}, where the embedding similarity is measured in terms of inner-product. With the shared candidate set, the siamese encoder, cross encoder and hybrid encoder will generate their recommendation results based on the ranking scores calculated with Eq.\ref{eq:siamese}, \ref{eq:cross} and \ref{eq:hybrid}, respectively.

$\bullet$ \textbf{System Settings}. All the models are implemented based on PyTorch 1.6.0 and Python 3.6.9, and trained on Nvidia V100 GPU clusters. The time cost is tested with a CPU machine (Intel(R) Xeon(R) CPU E5-2690 v4 @ 2.60GHz, with 112 GB MEM), which is consistent with the production environment\footnote{GPU inference is not feasible in production environment due to factors like huge online traffic, and the requirement of realtime service (user's request must be processed right away instead of being buffered and processed periodically)}. 

\subsection{Experiment Analysis}

\subsubsection{Basic Comparison}\label{sec:exp-ana-bas} In basic comparison, a total of 100 candidate ads are firstly generated via ANN search; then each encoder makes the top-N recommendation based on its computation of user-ad relevance. The experimental results are reported in Table \ref{tab:basic}, where both datasets and both types of backbone networks are included for comparison. It can be observed that hybrid encoder outperforms siamese encoder in all settings; and while used for native ads, it achieves comparable recommendation quality as the cross encoder. In Google News dataset, gaps between hybrid encoder and cross encoder are relatively larger, which may attribute to the following reason. The Google News dataset has much less training instances than the native ads dataset; as a result, the model's performance could be limited by insufficient training. However, the cross encoder can be more robust to this adverse situation because data can be utilized with higher efficiency. Specifically, as the input query and target are jointly encoded from the bottom-up, it becomes easier to figure out their underlying relevance; therefore, it enables the model to be learned more quickly and less affected by the data sparsity. To verify the above analysis, additional experiments are carried out in which models are trained with less data on native ads dataset (with 1/3 of the original training data). It is found that the cross encoder's performance becomes ``Hit@1:0.1054, Hit@3:0.2397'', while the hybrid encoder's performance becomes ``Hit@1:0.0891, Hit@3:0.2089'' (comprehensive results are included in Table \ref{tab:partial} in Appendix). In other words, the cross encoder is less affected, which further confirms our previous analysis. More discussion about both encoders' relationship will be made in \cref{sec:exp-ana-pro}.

\begin{table}[t]
\centering
\includegraphics[width=1.05\linewidth]{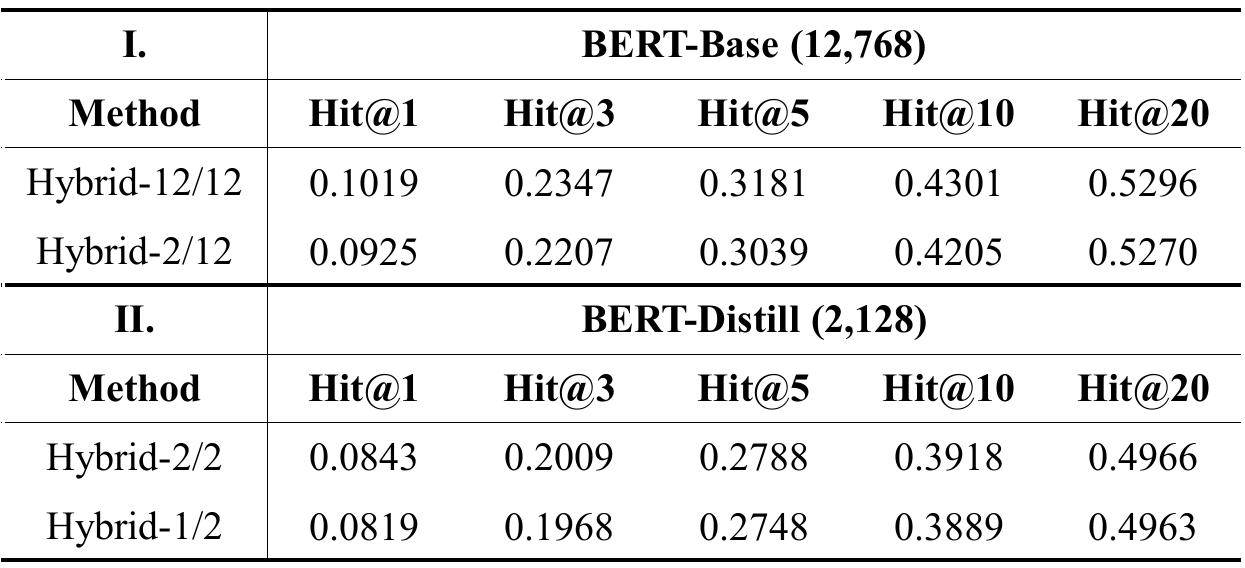}
\caption{Asymmetric Model Deployment \cref{sec:exp-ana-fr}. The light-weighted ua-interaction network leads to very limited loss in recommendation quality.}
\label{tab:asym}
\end{table}

\begin{table}[t]
\centering
\includegraphics[width=1.05\linewidth]{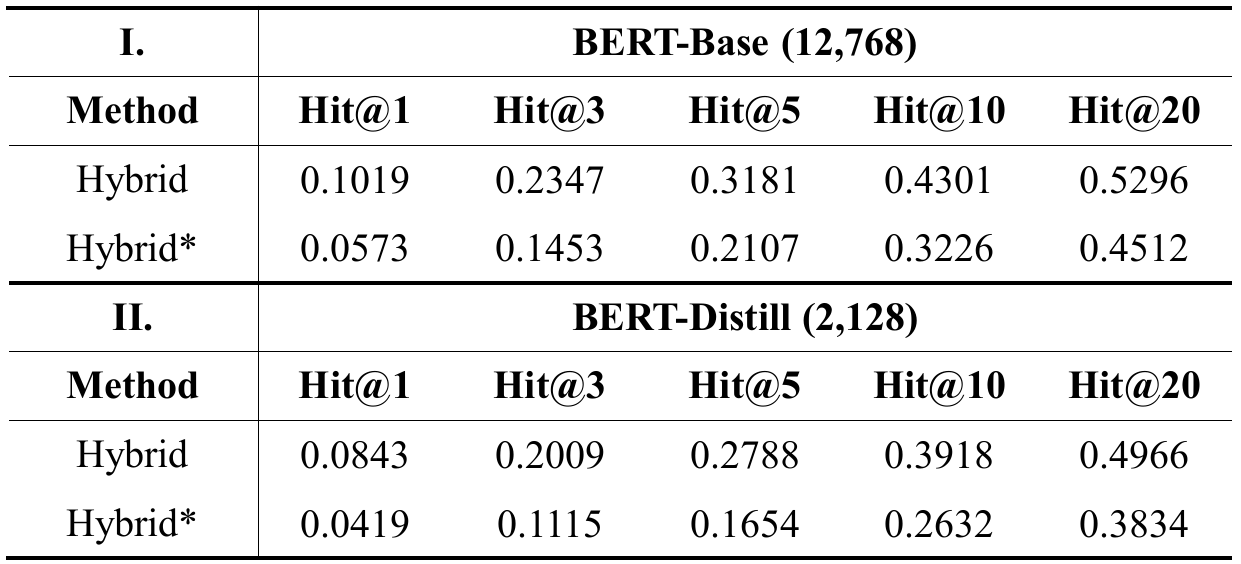}
\caption{Progressive Training \cref{sec:exp-ana-pro}. The progressive training significantly improves hybrid encoder's performance.}
\label{tab:progress}
\end{table}

\subsubsection{Hit-Time Ratio}\label{sec:exp-ana-tim} More detailed performance analysis is made for the model trained for real-world production (i.e., the one based on distilled BERT). Particularly, the top-1 recommendation quality is measured with variant number of candidate ads from ANN search, together with the inference time needed to process every single user. (In the previous experiment, it is set to be 100 regardless of time cost; but in production, it's required strictly to work in millisecond scale). As shown in Table \ref{tab:time}, the hybrid encoder merely incurs limited additional cost compared with the siamese encoder, meanwhile it improves the recommendation quality by a large-margin. As discussed in \cref{sec:met-ov-time}, the siamese encoder takes one transformer encoding to serve each user, while the hybrid encoder merely introduces another $\epsilon$ pass of feed-forward inference ($\epsilon=1$ for this experiment as ua-interaction network and user embedding network are of the same size). While in reality, the increment of running time is even humble as the inference of user embedding network and ua-interaction network are called simultaneously. In contrast, the cross encoder is far more expensive, as user needs to be encoded from scratch with each of the candidate ads. (Although in experiment environment, the encoding process can be run in parallel given a sufficient number of machines, it's not realistic in production as online traffic is tremendous). It can be observed that the model's running time grows rapidly and soon becomes over 10 ms, which is a default threshold for many realtime services, like native ads recommendation. Despite the expensive running cost, the performance gain from cross encoder is very limited: similar to the observation in \cref{sec:exp-ana-bas}, the hybrid encoder and cross encoder's performances are quite close to each other in most of the settings. 

\subsubsection{Asymmetric Model Deployment}\label{sec:exp-ana-fr} Recall that the hybrid encoder's working efficiency can be further improved if a smaller scale ua-interaction network is utilized (which makes $\epsilon$ lower than 1), we further test the model's performance with ua-interaction network and user/ad embedding network deployed asymmetrically. In particular, for user/ad embedding network with BERT-base (12-layer), the ua-interaction network is switch to a smaller 2-layer transformer; while for user/ad embedding network with distilled BERT (2-layer), the ua-interaction network is built on a single-layer transformer (the hidden-dimensions stay the same). The experiment result is demonstrated in Table \ref{tab:asym}. Despite that the ua-interaction network's scale is reduced significantly, the loss in quality is relatively small, especially when a large number of ads are recommended (Top-20). The underlying reasons are twofold. Firstly, the disentangled ad embeddings are pre-computed with the same network, whose quality remains the same. Secondly, although the ua-interaction network is simplified, user's information for analyzing user-ad relevance may still be effectively extracted thanks to its being attended by the ad. 

\subsubsection{Progressive Learning}\label{sec:exp-ana-pro} The ablation study is made to evaluate the progressive training's effect on native ads recommendation \cref{sec:met-pro}. The experiment results are demonstrated in Table \ref{tab:progress}, where * means that the model is trained only with the global contrast learning. It can be observed that the hybrid encoder's performance is improved significantly when the progressive training is utilized. We further compare the progressive training pipeline's effect on cross encoder, and find that the cross encoder's performance may also benefit from it; e.g., in native ads dataset with BERT-base backbone, the performance is improved from ``Hit@1:0965, Hit@3:0.2178'' to ``Hit@1:0.1111, Hit@3:0.2497'' (more results are included in Table \ref{tab:progress-a} in Appendix). However, it is not as significant as hybrid encoder, and such a phenomenon may still be resulted from the cross-encoder's higher data efficiency in learning user-ad relevance, as discussed in \cref{sec:exp-ana-bas}. These observations do not conflict with our basic argument; in fact, the cross encoder is always more capable of making fine-grained prediction, while the hybrid encoder may get very close to the cross encoder's performance if it is trained properly and sufficiently.

\subsubsection{Degree of Disentanglement}\label{sec:exp-ana-deg} The disentanglement degree's impact on native ads recommendation is also studied (i.e., the number of disentangled ad embeddings to be generated). The experiment results are demonstrated in Table \ref{tab:degree}, where hybrid$^i$ means that the degree is set to be $i$. It can be observed that the best performance is achieved at degree 1 and 3, when the backbone structure is BERT-base (12-layer-768-dim) and distilled BERT (2-layer-128-dim), respectively. Meanwhile, further increasing the disentanglement degree will not help the model's performance. Such a phenomenon indicates that a large-scale model (e.g., BERT-base) may fully represent ad's information with one-single high-dimension vector; and when the model's scale is reduced (e.g., distilled BERT), it may require a few more vectors to fully extract ad's information. Because of the simplicity about ad's description (usually a short sequence of keywords indicating one particular product or service), a small disentanglement degree will probably be sufficient in practice.

\subsubsection{Summarization} The major findings of the experimental studies are summarized as follows.
\\
$\bullet$ The hybrid encoder significantly outperforms the siamese encoder with little additional cost; besides, it achieves comparable recommendation quality as the cross encoder, but works with much higher efficiency.
\\
$\bullet$ The progressive training pipeline contributes substantially to the model's performance. Because of the simplicity of ad description, a small disentanglement degree can be sufficient in practice. Besides, a largely simplified ua-interaction network is able to reduce the model's inference cost with very limited loss in quality. 

\begin{table}[t]
\centering
\includegraphics[width=1.05\linewidth]{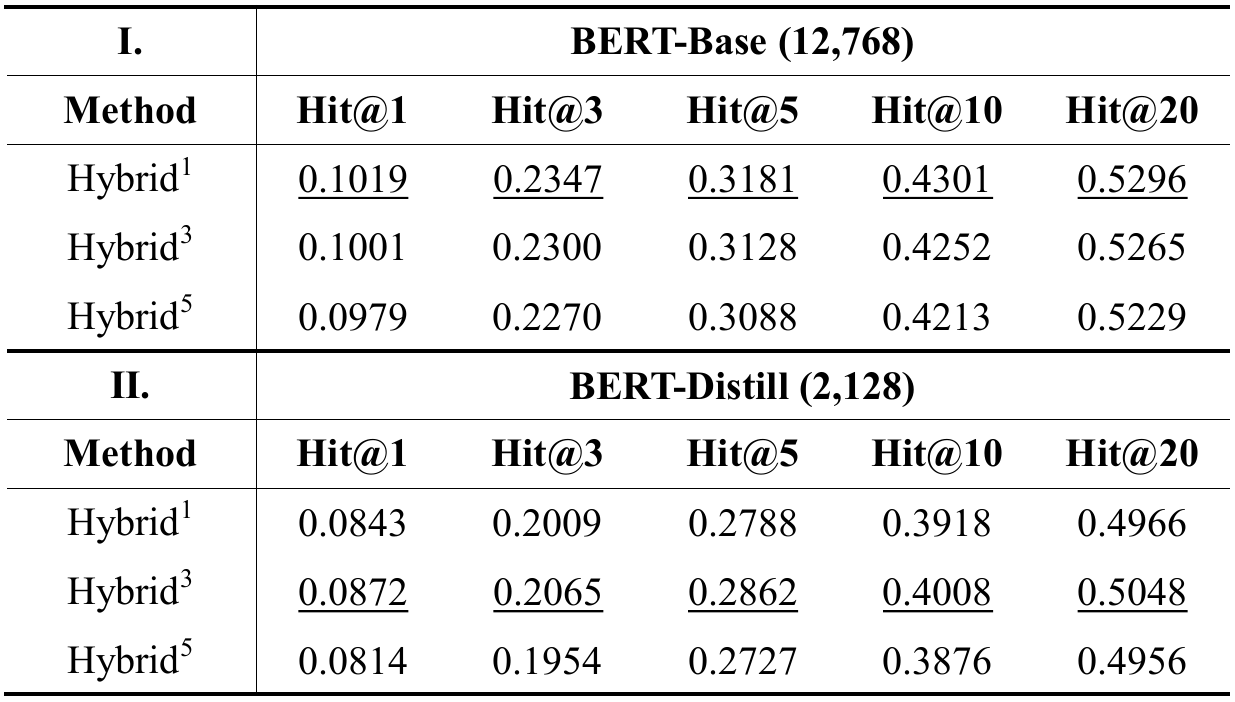}
\caption{Degree of Disentanglement on Native Ads \cref{sec:exp-ana-deg}.}
\label{tab:degree}
\end{table}

\section{Conclusion}
In this paper, we proposed hybrid encoder for the efficient and precise native ads recommendation. 
With hybrid encoders, the recommendation is made through two consecutive steps: retrieval and ranking. 
In the retrieval step, user and ad are independently encoded, based on which candidate ads can be acquired via ANN search. 
In the ranking step, user and ad are collaboratively encoded, where ad-related user embeddings and disentangled ad embeddings are generated to select high-quality ads from the given candidates.
Both the retrieval step and ranking step are light-weighted, which makes hybrid encoder competitive in working efficiency. Besides, a progressive training pipeline is proposed to optimize the overall performance of the two-stage workflow, where the model is learned to learned to ``make candidate retrieval from the global scope'' and ``select the best ads from the given candidates'' step-by-step. The experimental studies demonstrate that hybrid encoder significantly improves the quality of native ads recommendation compared with conventional siamese encoder; meanwhile, it achieves comparable recommendation quality as the cross encoder with highly reduced time consumption.

\bibliographystyle{Format/ACM-Reference-Format}
\bibliography{ref}

\appendix
\section{Appendix}
Additional experimental results (shown in the next page) are included in the Appendix. According to Table \ref{tab:progress-a}, the models' performances are consistently improved with the adoption of progressive training. And in Table \ref{tab:partial}: cross encoder's performance is less affected when the volume of training data is reduced.

\newpage

\begin{table*}[t]
\centering
\includegraphics[width=1.0\textwidth]{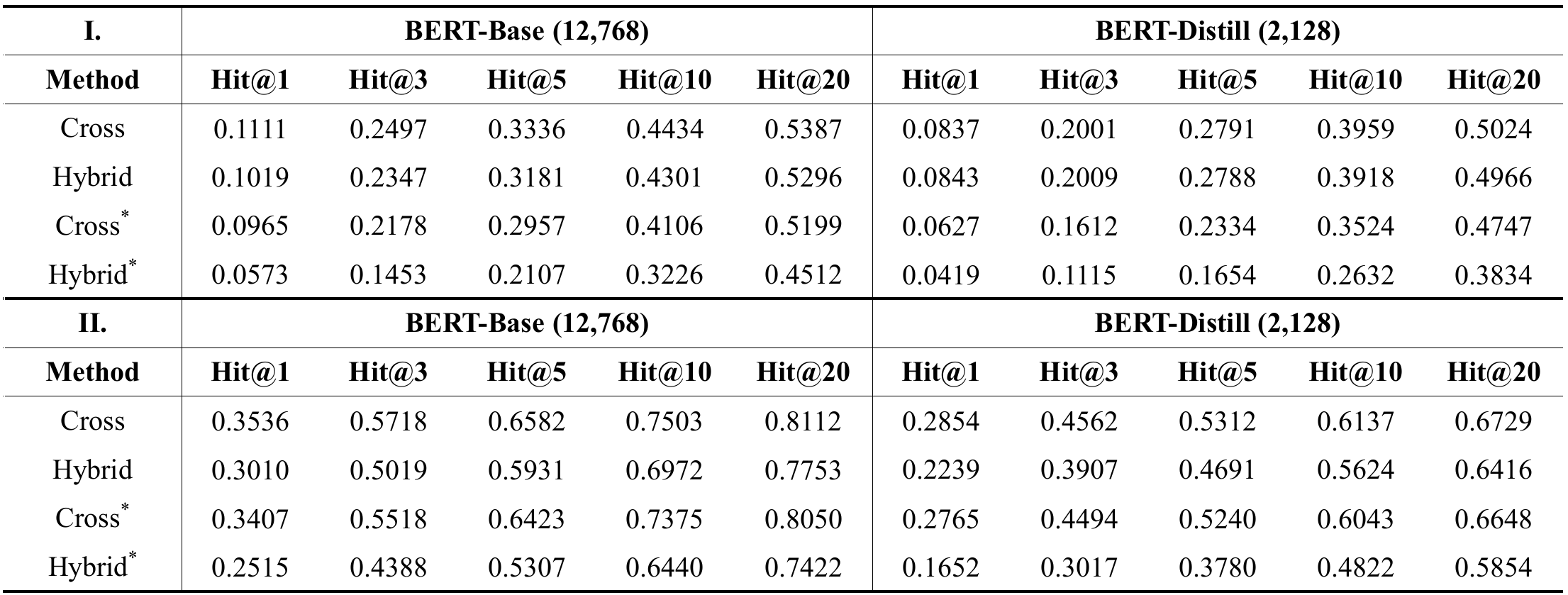}
\caption{The comprehensive result of hybrid encoder w./w.o. progressive training pipeline. (I.: Native Ads, II: Google News)}
\label{tab:progress-a}
\end{table*}

\begin{table*}[t]
\centering
\includegraphics[width=0.55\textwidth]{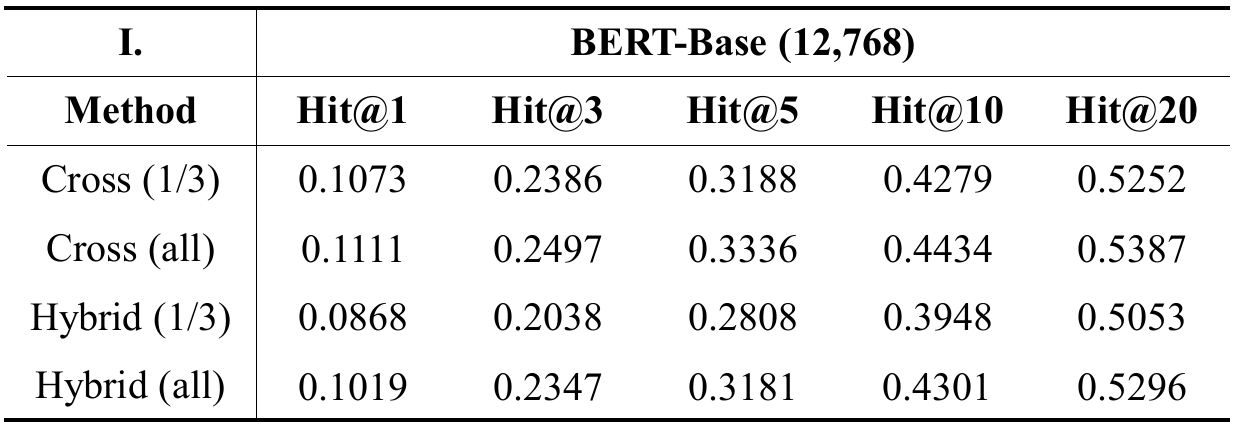}
\caption{Hybrid Encoder's performance with 1/3 of the original training data.}
\label{tab:partial}
\end{table*}

\end{document}